\documentclass[letterpaper]{article} % DO NOT CHANGE THIS
\usepackage{aaai2027}  % DO NOT CHANGE THIS
\nocopyright % arXiv preprint: suppress the AAAI copyright notice
\usepackage[hyphens]{url}  % DO NOT CHANGE THIS
\usepackage{graphicx} % DO NOT CHANGE THIS
\usepackage{natbib}  % DO NOT CHANGE THIS AND DO NOT ADD ANY OPTIONS TO IT
\usepackage{caption} % DO NOT CHANGE THIS AND DO NOT ADD ANY OPTIONS TO IT
\usepackage{booktabs}

\usepackage{amsmath}
\usepackage{amssymb}
\usepackage{amsfonts}
\usepackage{multirow}
\usepackage{xspace}
\newcommand{\sysname}{IDP AutoOpt\xspace}

\definecolor{keyorange}{rgb}{0.60,0.39,0.00} % orange!60!black, 5.0:1
\definecolor{keyred}{rgb}{0.70,0.00,0.00}    % red!70!black,    5.9:1

\title{IDP AutoOpt: Agent-Driven Optimization of Document Processing Pipeline Configurations}

\author{
    David Kaleko,
    Sergey Ivanov,
    Md Mofijul Islam
}
\affiliations{
    Amazon Web Services\\
    kaleko@amazon.com, ssivanov@amazon.fr, miislamg@amazon.com
}

\begin{document}

\maketitle

\begin{abstract}
We present \sysname, an autonomous LLM agent that discovers high-performing configurations for intelligent document processing (IDP) pipelines. Tuning IDP prompts, models, OCR settings, and schemas jointly currently costs domain specialists 20 to 80+ person-hours per document type and does not scale as enterprises add document classes. \sysname runs a closed loop: it scores a configuration on a small labeled set, diagnoses field-level errors, generates targeted edits, and re-evaluates, guided by human-authored domain skills that encode production expertise. Across extraction, classification, and packet-splitting tasks deployed in healthcare, marketing-intelligence, and financial-services settings, \sysname matches or exceeds human-expert accuracy at equal or lower cost (on an extraction benchmark, $88.0\%$ vs $81.6\%$ at $4.6\times$ lower per-page cost), cutting configuration time from weeks to under two hours. We further show a sharp capability gap among the agent LLMs we tested, below which optimization fails outright, and that curated domain skills tend to outperform raw source-code access as a means of supplying domain expertise to the agent. We also share practical lessons on context management and variance mitigation. Requiring only a configurable pipeline, a scoring function, and a small labeled set, we expect the approach to extend beyond IDP to other enterprise AI systems, such as RAG and multi-agent workflows, where configuration bottlenecks deployment.
\end{abstract}

% Uncomment the following to link to your code, datasets, an extended version or similar.
% You must keep this block between (not within) the abstract and the main body of the paper.
% Make sure that you do not de-anonymize yourself with these links.
% \begin{links}
%     \link{Datasets}{https://huggingface.co/datasets/amazon-agi/RealKIE-FCC-Verified}
% \end{links}

% =====================================================================
% Main body. One file per section under sections/ so sections can be
% edited independently.
%
% NOTE FOR SUBMISSION: AAAI requires the source to be a SINGLE .tex file
% ("do not use the input command to include sections of your paper ---
% every section must be in the single source file"; only the .bib may be
% separate). Run `python3 flatten.py` to expand every \input below into
% main-flat.tex, and submit that file instead of this one.
% =====================================================================
\section{Introduction}

Modern intelligent document processing (IDP) systems extract structured information from unstructured documents by orchestrating OCR engines, large language models, classification logic, and post-processing rules into configurable pipelines \cite{yoon-etal-2024-lofi, wang2025hybridocr, islam2026docsplit}. We study this recent generation of pipelines rather than the classical, template-and-rule variety: LLMs now carry every stage, reading the page directly or on top of an OCR backend, deciding the document class, extracting fields against a schema, splitting multi-document packets, and estimating confidence in their own outputs. These systems have become critical infrastructure across industries, with deployments processing hundreds of thousands of documents daily. Their components have improved rapidly, from layout-aware pretraining \cite{xu2020layoutlm} and OCR-free encoders \cite{kim2022donut} to multimodal models that read a rendered page directly \cite{mathew2021docvqa, shi2024docparsing}. In production, however, end-to-end accuracy is rarely limited by the components themselves. It is limited by how they are assembled and configured.

Achieving high accuracy requires extensive configuration: document schemas, OCR settings, model selection, document classification, extraction prompts strategies must all be tuned jointly. This configuration space is fundamentally heterogeneous, spanning natural language, categorical choices, continuous parameters, structural decisions, and semi-structured data. Unlike classical hyperparameter tuning \cite{lin2025storagextuner, zhou2023dbot}, IDP configuration requires the kind of reasoning a human engineer performs: diagnosing failure patterns, rewriting prompts based on observed errors, and making structural pipeline changes.

Today that reasoning is supplied entirely by hand. Across the production engagements behind this work, domain specialists invest 20 to 80+ person-hours per document type before a configuration reaches its accuracy target, and one customer maintains 80+ distinct document configurations in a single deployment (Section~\ref{sec:deployment}). The cost is linear in the number of document types and it recurs on every model deprecation, price change, schema revision, and new layout, so the specialists who understand the pipeline best spend their time re-tuning it rather than extending it. Manual configuration, not model quality, is what currently bounds how widely IDP can be deployed.

\begin{figure*}[t]
\centering
% The vector source of this diagram embedded CID/Identity-H TrueType
% subsets and a Type 3 NotoColorEmoji font. AAAI rejects both, even
% inside a graphic, so the figure ships as a flattened 400 dpi raster
% (727 dpi at this printed width) with no fonts of any kind.
\includegraphics[width=0.9\textwidth]{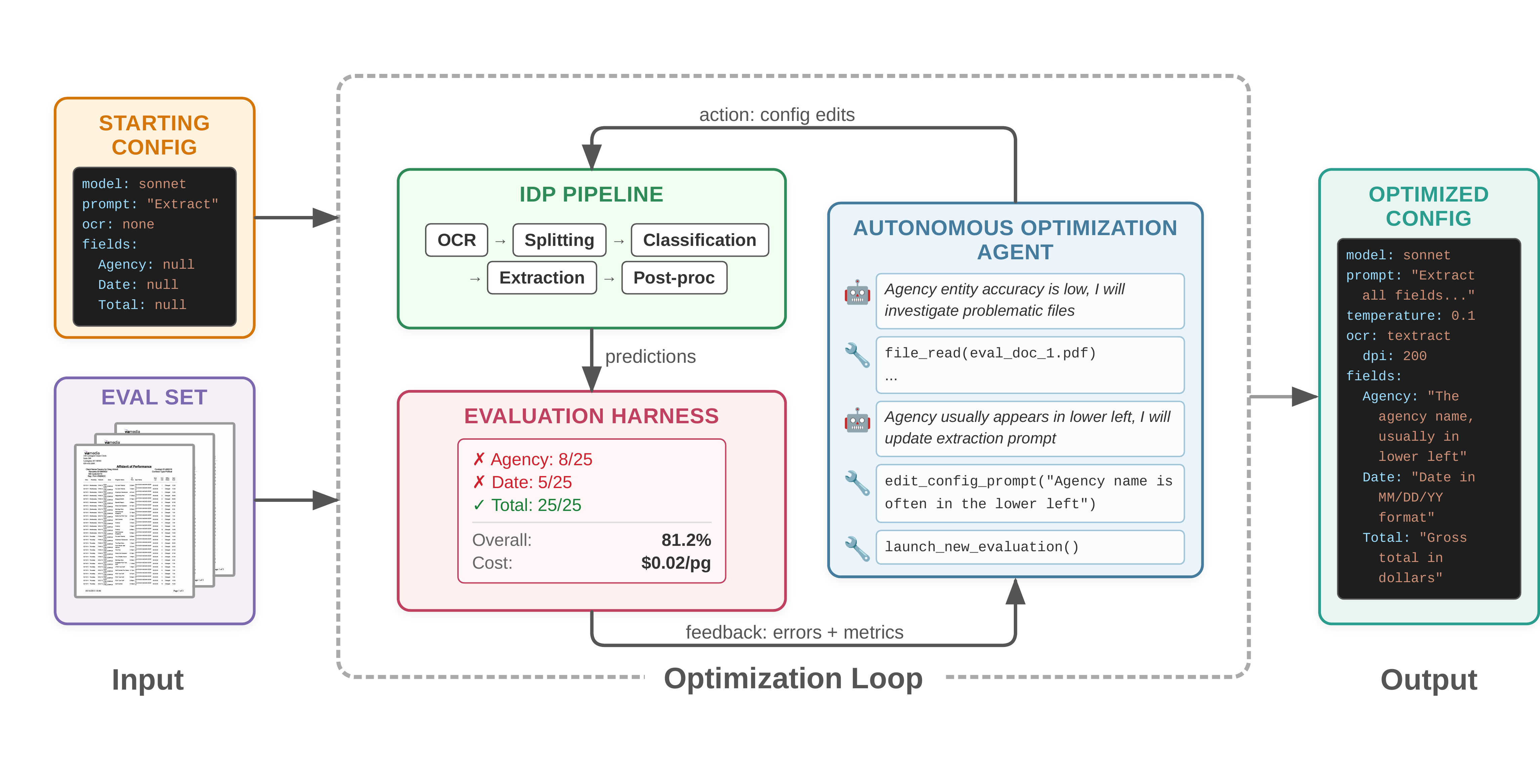}
\caption{\sysname optimization loop. The system takes a \textcolor{keyorange}{starting configuration} (YAML) and a \textcolor{violet}{labeled evaluation set} as input. Inside the loop, the \textcolor{green!50!black}{IDP pipeline} produces predictions that the \textcolor{keyred}{evaluation harness} scores per-field. The \textcolor{blue}{autonomous optimization agent} diagnoses errors and generates targeted configuration edits. After iterating, the system outputs an \textcolor{teal}{optimized configuration}.}
\label{fig:system_overview}
\end{figure*}

Automating this tuning is harder than it first appears, for four reasons. \textbf{(1)} \textit{The search space is heterogeneous and composable.} Classical hyperparameter and AutoML search assumes a fixed-dimensional space of numeric and categorical knobs over which a surrogate model can be fit \cite{bergstra2012randomsearch, snoek2012practicalbo, feurer2015autosklearn}; prompt optimizers search free-form text while holding pipeline structure fixed \cite{zhou2023ape, pryzant2023apo, yang2024opro, wang2024promptagent, opsahlong2024mipro}; and compound-system frameworks optimize modules inside a program the developer has already declared \cite{khattab2024dspy, yuksekgonul2024textgrad}. An LLM-based IDP configuration contains all of these at once, and because the same model is prompted at several stages, its dimensions interact across them: enabling layout-aware OCR changes which prompt wording helps downstream, and a schema edit changes which few-shot example is informative. \textbf{(2)} \textit{Feedback is expensive, coarse, and noisy.} Scoring one candidate means running the full pipeline over the labeled set, which costs real money and minutes of wall-clock time, so the realistic budget is tens of evaluations rather than the thousands that random or Bayesian search assumes. Labeled data is itself the scarce resource that motivates automation, so evaluation sets are small, which raises the risk of fitting them rather than the task. \textbf{(3)} \textit{Diagnosis is a reasoning problem, not a sampling problem.} A field-level accuracy score says which fields failed but not why: a checkbox field that text-only OCR cannot see, a date rendered as a bitmask, and a prompt that never mentions a field are three different failures behind the same number, and separating them requires inspecting the document itself alongside the schema and the prompt. \textbf{(4)} \textit{Long-horizon autonomy is fragile.} A multi-hour run generates far more tokens than any context window holds, and model performance degrades as relevant evidence is buried in long contexts \cite{liu2024lostinmiddle}; agents self-correct unreliably without external verification \cite{huang2024selfcorrect, shinn2023reflexion, madaan2023selfrefine}; and an agent rewarded by a scalar score will exploit whatever shortcut raises it \cite{skalse2022rewardhacking, pan2022rewardmisspec}, a failure we observed directly and report in Section~\ref{sec:generalization}. Together these make the credibility of any reported improvement contingent on how the experiment is run, not only on the system.

Recent work has shown that LLM agents can optimize software through closed-loop iteration \cite{zhou2023dbot, lin2025storagextuner, chi2024sela, yang2024opro, wang2024promptagent, yuksekgonul2024textgrad}, building on the observation that tool-using agents can interleave reasoning with acting \cite{yao2023react} and act competently on real engineering artifacts \cite{jimenez2024swebench}. However, none of these systems address the challenges of IDP configuration, where the search space is broader, composable and more heterogeneous than numeric knobs, code, or prompt text alone.

To address this, we developed \sysname, an autonomous LLM agent that automates IDP configuration through iterative evaluation-driven optimization. Given a small labeled set of documents and a minimal starting configuration, the agent operates in a closed loop: analyzing errors, generating targeted modifications, evaluating, and iterating, guided by domain skills encoding production expertise. Our contributions are: \textbf{(1)} a formulation of IDP tuning as constrained optimization over a heterogeneous configuration space, together with an agent architecture that searches it directly, editing prompts, schemas, model choices, OCR settings, and pipeline structure in a single loop under an explicit per-page cost constraint (Section~\ref{sec:system}); \textbf{(2)} human-authored \textit{domain skills}, structured modules that encode the diagnostic heuristics production specialists actually use, as the mechanism for injecting expertise without retraining; \textbf{(3)} an evaluation protocol built for this setting, with a test set the optimizer cannot reach by construction, configuration selection driven by evaluation accuracy alone, and a noise-aware threshold for credible improvement derived from repeated runs (Sections~\ref{sec:setup} and~\ref{sec:results}); and \textbf{(4)} an empirical study spanning 44 optimization runs across five agent LLMs, four knowledge conditions, and three IDP tasks, alongside production deployments in healthcare, marketing-intelligence, and financial-services settings.

Our experiments yield four findings. First, \sysname exceeds human expert accuracy while cutting cost: starting from a minimal configuration at roughly 70\% accuracy, it passes a human-expert baseline built over weeks of manual tuning within four iterations, and the configuration a user would actually select reaches 88.0\% held-out accuracy at \$0.022 per page, 6.4 points above the expert at 4.6$\times$ lower inference cost, in under 2 hours of autonomous operation. Second, the approach generalizes across IDP tasks and into production: the agent matches expert accuracy on 9-class classification at lower cost and improves packet splitting from 75.9\% to 84.1\% at identical cost, and in a deployment processing tens of millions of pages per month it beat a specialist team's configuration at 3.7$\times$ lower cost per page. Third, agent capability is the binding constraint: a sharp gap separates the LLMs we tested, with two of five never improving on the starting configuration while the other three produce gains well outside run-to-run noise, and the failures are outright rather than merely slower. Fourth, curated domain skills tend to outperform raw source-code access, which ranks last for both models we ablated because it diverts the agent's budget toward orientation instead of targeted edits. We additionally report practical deployment considerations, including convergence behavior, run-to-run variance, context management, and a reward-hacking incident that shaped our protocol. Throughout, we report results under a strict eval/test selection protocol (Section~\ref{sec:setup}) and assess accuracy comparisons against a noise-aware threshold derived from repeated runs (Section~\ref{sec:results}).

Nothing in the loop is specific to documents. \sysname assumes only a configurable pipeline, a scoring function, and a small labeled set, so the same design should extend to other configuration-bottlenecked systems such as RAG and multi-agent workflows.

\section{\sysname: System Design}
\label{sec:system}

We frame the task as constrained optimization over a compound processing pipeline. Consider a multi-component system $P$ (in our case, a document processing pipeline combining OCR, classification, and extraction stages) whose behavior is governed by a configuration $c$ (a YAML file specifying pipeline setup details, prompts, model choices, schemas, and parameters). Given a small labeled evaluation set $D$ and a scoring function $S$ that measures how well the pipeline's output matches ground truth, we seek:

\begin{equation}
    c^* = \arg\max_{c \in \mathcal{C}} S(P(c), D) \quad , \quad c \in \Omega
\end{equation}
where $\Omega$ encodes operational constraints such as cost budgets, latency requirements, or model availability. In our IDP instantiation, $S$ computes field-level extraction accuracy (how many fields were extracted correctly), and the primary constraint $\Omega$ imposes a per-page inference cost ceiling that reflects real customer budgets.

\subsection{Architecture and Optimization Loop}

\sysname consists of three components (Figure~\ref{fig:system_overview}): \textbf{(1)} an \textit{IDP pipeline}, the document processing platform invoked programmatically through an evaluation API; \textbf{(2)} an \textit{evaluation harness} that computes field-level accuracy, generates per-document error breakdowns, and versions every configuration the agent produces; and \textbf{(3)} an \textit{autonomous optimization agent}, a multimodal LLM with tool access that reasons over evaluation metrics, inspects document images, and produces configuration edits without any human intervention.

The optimization loop proceeds iteratively. At each step, the agent receives feedback describing which fields failed, on which documents, and with what patterns. It diagnoses root causes, sometimes inspecting document images via multimodal vision (e.g., to understand why a checkbox field fails when text-only OCR cannot detect it), then generates targeted edits: rewriting a prompt, adding a field description, switching an OCR backend, or inserting a few-shot example. The modified configuration is evaluated over the full labeled set, and the results inform the next decision. Throughout, the agent maintains an append-only \textit{optimization log} recording every decision, rationale, and outcome, which serves as both long-term memory and a human-readable audit trail.

The agent terminates at a fixed iteration budget (50), a wall-clock timeout (8 hours), or when optimization cost exceeds a limit. Optionally, it has read-only access to the target pipeline's source code, a realistic scenario for teams optimizing their own systems (impact measured in Section~\ref{sec:results}).

\subsection{Configuration Space}
\label{sec:config_space}

The YAML configuration governs a multi-stage processing pipeline, each with independent settings. The agent can modify: \textbf{(1)} natural-language system/task prompts, JSON Schema field definitions, and few-shot examples; \textbf{(2)} per-stage LLM choice and hyperparameters (temperature, max tokens, top-$p$); \textbf{(3)} pipeline structure (extraction strategy, classification method, splitting logic, assessment granularity); \textbf{(4)} OCR backend, feature toggles (layout analysis, table detection), and image resolution; and \textbf{(5)} operational knobs (confidence thresholds, prompt caching, regex bypasses, feature flags). This space simultaneously includes natural language, categorical variables, continuous parameters, structural choices, and semi-structured data, which distinguishes it from tuning numeric knobs or optimizing prompts in isolation.

\subsection{Domain Skills}
\label{sec:skills}

Domain skills are structured knowledge modules encoding expert heuristics from production engagements, each specifying a trigger condition, diagnostic steps, and resolution strategy. The system includes 27 skills distilled from 8+ engagements. Appendix~\ref{sec:app_skills} gives three worked examples. Unlike auto-generated approaches \cite{chen2026skillforge, wang2026skillos}, our skills are human-authored from experience, trading scalability for reliability. The authoring cost is modest: a domain expert starting cold takes roughly an hour per skill, and in practice most skills are a byproduct of engagement work already being performed, since the diagnostic reasoning they encode is what the expert did to resolve the customer's issue in the first place. The 27 skills used here were written incrementally across 8+ engagements rather than as a dedicated authoring effort. Whether this investment outperforms zero-effort alternatives is tested in Section~\ref{sec:results}. A complete example skill is provided in Appendix~\ref{sec:example_skills}.

\subsection{Context Management}
\label{sec:context}

Long optimization runs exceed any model's context window. \sysname uses proactive compaction: when token usage exceeds a threshold (\% of context window), older history is summarized while the optimization log is re-injected to preserve decision continuity. Details are in Appendix~\ref{sec:context_detail}.

\section{Real-World Deployment}
\label{sec:deployment}

\sysname is deployed in production and has been applied to customer engagements spanning healthcare document management, marketing intelligence, and financial services. In this section, we describe the deployment context that motivates our system and the production results that validate it.

\paragraph{The configuration bottleneck:} Based on data from multiple engagements, domain specialists typically invest 20 to 80+ person-hours per document type to achieve target accuracy. Complex multi-class deployments (one customer manages 80+ document configurations) require several weeks of iterative refinement. As organizations scale IDP to more document types and higher volumes, manual per-type tuning becomes unsustainable.

\paragraph{Production results:} In one deployment processing tens of millions of pages per month, a team of specialists spent several weeks creating an extraction configuration that achieved 88\% accuracy. Starting from scratch, \sysname produced a configuration reaching 89.5\% accuracy at 3.7$\times$ lower cost per page in a few hours of autonomous operation. On a separate public benchmark, the agent improved extraction accuracy from 75.5\% to 89.3\% without model changes or cost increases (see Section~\ref{sec:generalization}).

\section{Experimental Setup}
\label{sec:setup}

\paragraph{Dataset:}
We evaluate on the RealKIE FCC-Verified dataset \cite{amazonagi2024realkiefccverified}, a collection of 75 real-world broadcast advertising invoices spanning 18 distinct visual layouts from different TV stations and media companies. The dataset is split into an \textit{eval set} of 25 documents (manually selected to represent all layout templates) and a held-out \textit{test set} of 49 documents (one excluded due to annotation quality issues). The agent optimizes exclusively on the eval set; the test set is reserved for final evaluation of selected configurations. This relatively small dataset was chosen because it reflects a common user pain point of generating extensive high quality labeled data.

\paragraph{Test-set isolation protocol:} The test set is inaccessible to the optimizer by construction, not merely by convention: test documents are never uploaded to the AWS account in which the agent runs, every configuration the agent selects is based solely on eval-set accuracy, and test-set inference is performed post hoc in a separate account. No reported number is obtained by maximizing over test performance. Appendix~\ref{sec:app_isolation} states the full protocol.

\paragraph{Task and metrics:}
The task is key information extraction: given an invoice document, extract 7 top-level fields (Agency, Advertiser, GrossTotal, PaymentTerms, AgencyCommission, NetAmountDue) and a variable-length array of line items (each with 5 sub-fields). All fields use strict exact-match evaluation; strings must match character-for-character and numbers must match exactly. We report two metrics, both of which select configurations using eval accuracy only. (1) \textit{Peak test accuracy} is the held-out test accuracy \textit{of the configuration that scored highest on the eval set} while respecting the cost constraint. It is deliberately not the maximum of the test curve: taking that maximum would require the test set to inform selection, which our protocol forbids. (2) \textit{AUC-50-Calibrated} is the area under the test accuracy curve over 50 iterations, baseline-subtracted (zero means no improvement), where the sequence of configurations is the running maximum \textit{of eval accuracy} and each such configuration is then scored on test. Both metrics therefore answer ``how well do the configurations a user would actually have chosen generalize?'' rather than ``what is the best test score obtainable in hindsight?''

\paragraph{Initial configuration and human baseline:}
All experiments begin from the same minimal configuration: field names with null descriptions, a generic prompt, no few-shot examples, no OCR, and Claude Sonnet 4.6 as extraction model ($\sim$70\% test accuracy, zero human effort). As a comparison, Table~\ref{tab:human_baseline} shows domain specialists manually tuning the same dataset over several weeks, reaching 81.6\% at \$0.102/page. The comparison is matched on access --- same documents, models, OCR backends, configuration space, and scoring harness --- with one intentional asymmetry: the agent was capped at \$0.05/page while the human baseline was uncapped and its final configuration costs \$0.102/page. Appendix~\ref{sec:app_parity} details the protocol.

\begin{table}[t]
\centering
\small
\begin{tabular}{lrrl}
\toprule
\textbf{Version} & \textbf{Acc.} & \textbf{\$/page} & \textbf{Description} \\
\midrule
V1 & 76.2\% & 0.102 & Haiku + Textract full \\
V3 & 78.3\% & 0.027 & Nova 2 Lite + layout OCR \\
V5 & 75.2\% & 0.028 & Simplified prompt \\
V\_current & 81.6\% & 0.102 & Rich descriptions \\
\bottomrule
\end{tabular}
\caption{Human expert configuration trajectory on RealKIE test set. Each version represents days to weeks of iterative manual tuning. Weeks of effort plateau at 81.6\% accuracy.}
\label{tab:human_baseline}
\end{table}

\paragraph{Ablation conditions:}
We vary both the agent LLM (Claude Opus 4.7, Claude Sonnet 4.6, Claude Haiku 4.5, Kimi K2.5, Llama 4 Maverick) and the domain knowledge available for the agent to read (skills + source code, skills only, source code only, neither). Each condition is repeated 2--3 times with identical starting conditions to measure run-to-run variance. All runs use 50 iterations, an 8-hour timeout, and a cost constraint of \$0.05/page.

\section{Experimental Results}
\label{sec:results}

\begin{figure*}[t]
\centering
% TODO(figure): this plot needs to be regenerated before submission.
% (a) Resolution: the source raster is only 1500x900 px. AAAI requires
%     300 dpi at the printed size, which caps the width at ~5in
%     (0.7\textwidth = 306 dpi); at full \textwidth it is 214 dpi and
%     non-compliant. Re-export at >=2100 px wide for full-width use.
% (b) Label size: AAAI requires text inside an illustration to be at
%     least 9 pt. The tick labels and in-plot annotations currently
%     print at roughly 3.4 pt here (~4.8 pt even at full \textwidth),
%     so the plot must be re-rendered with larger fonts and probably
%     fewer in-plot annotations. This cannot be fixed by rescaling.
\includegraphics[width=0.7\textwidth]{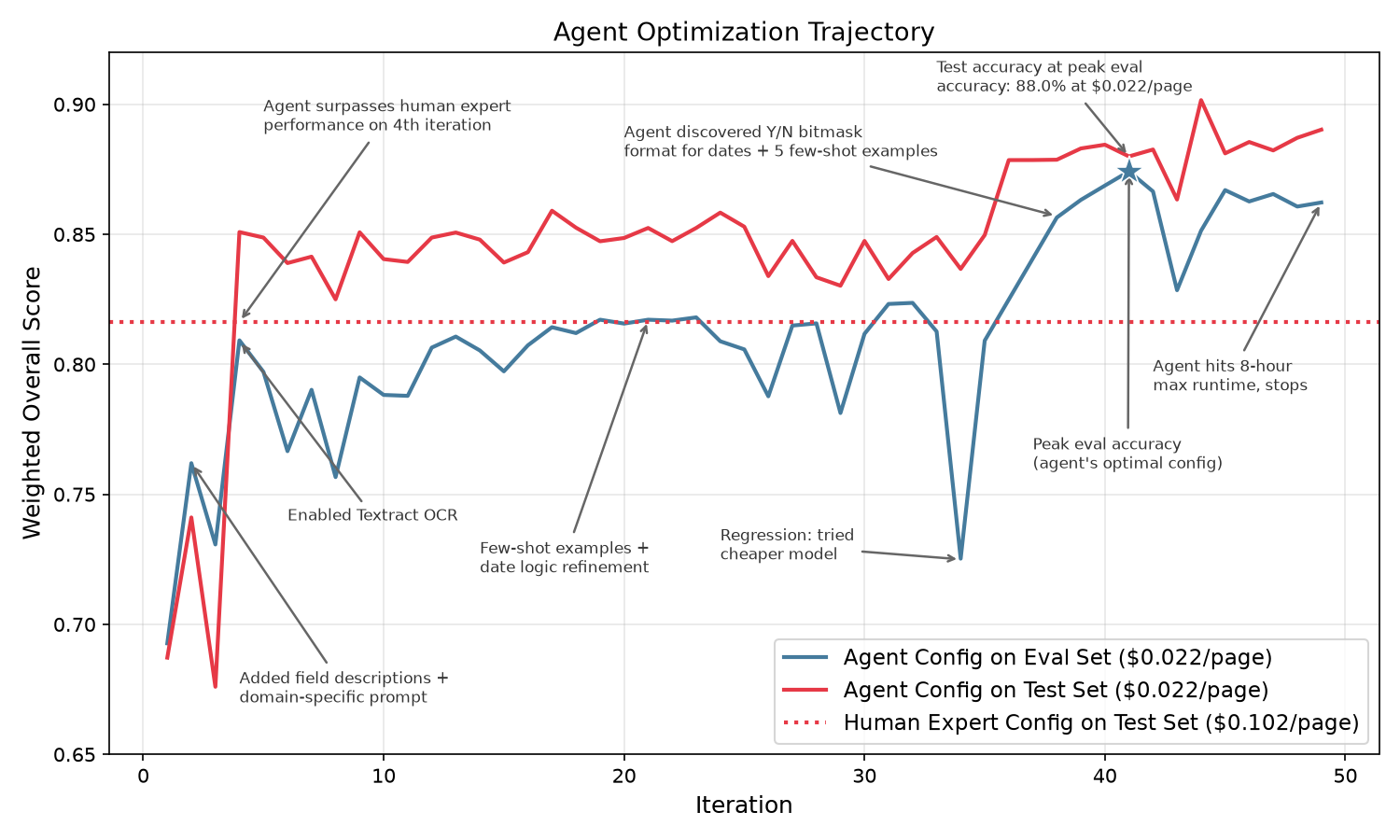}
\caption{\sysname optimization trajectory (Sonnet 4.6, skills + source code). \textcolor{blue}{Blue}: eval set accuracy. \textcolor{keyred}{Red}: held-out test set accuracy. \textcolor{keyred}{Dotted}: human expert baseline on held-out test set (81.6\%, \$0.102/page). The agent surpasses the expert by iteration 4. At the point of maximum eval accuracy (iteration 41, marked $\star$; the configuration a user would actually select), test accuracy is 88.0\% at \$0.022/page: +6.4 pp accuracy at 4.6$\times$ lower cost. The two curves rise and fall together, with no systematic divergence. Late iterations continue improving with no saturation at iteration 50, suggesting longer runs could push accuracy further.}
\label{fig:convergence}
\end{figure*}

\subsection{Comparison Against Human Experts}

Figure~\ref{fig:convergence} shows the trajectory of a single Sonnet 4.6 optimization run. Starting from a minimal 70\% configuration, the agent surpasses the human expert baseline (81.6\%, produced over weeks of manual tuning) by its 4th iteration, after adding robust field descriptions and enabling OCR. It continues improving through prompt refinement, few-shot examples, and date format fixes (iterations 5--20), then makes increasingly targeted discoveries such as a Y/N bitmask encoding for date-type fields (iterations 20--50). Eval accuracy is highest at iteration 41; that configuration scores 88.0\% on the held-out test set at \$0.022/page, which is 6.4 percentage points above the human expert at 4.6$\times$ lower cost per page. We stress that 88.0\% is the test score of the eval-selected configuration, not the maximum of the test curve. Individual later iterations do reach higher test accuracy, but selecting them would require test-set access the agent does not have, so we do not report them.

The agent also makes mistakes. At iteration 34, it experiments with a cheaper model that causes a regression, but detects the accuracy drop and reverts to the prior configuration. This ability to explore risky changes and recover is a structural advantage of versioned configuration history. Notably, the test accuracy (\textcolor{keyred}{red} in Figure~\ref{fig:convergence}) tracks the eval accuracy (\textcolor{blue}{blue}) throughout, with no systematic divergence, showing no evidence of overfitting when optimizing on 25 documents.

\noindent\fbox{\parbox{0.95\columnwidth}{\textbf{Finding 1.} \sysname exceeds the accuracy of human expert configurations (+6.4 pp) at 4.6$\times$ lower inference cost, reducing configuration time from weeks of specialist effort to under 2 hours of autonomous operation.}}

\paragraph{Cost accounting.} The 4.6$\times$ figure compares \textit{final per-page inference cost}, computed with the same formula for the human and agent configurations (\$0.102/page vs.\ \$0.022/page). It excludes the one-time cost of optimization itself, which was \$139.47 for the run in Figure~\ref{fig:convergence} and is amortized over the deployment's lifetime; Appendix~\ref{sec:app_cost} expands on this.

\subsection{Effect of Agent LLM Choice}

\begin{table}[t]
\centering
\small
\begin{tabular}{lrrc}
\toprule
\textbf{Agent LLM} & \textbf{AUC-50} & \textbf{Peak Acc.} & \textbf{\$/page} \\
\midrule
Claude Sonnet 4.6 & 5.28 & 88.5\% & 0.008 \\
Claude Opus 4.7 & 3.89 & 84.9\% & 0.019 \\
Kimi K2.5 & 3.72 & 84.1\% & 0.016 \\
Claude Haiku 4.5 & 0.38 & 77.6\% & 0.034 \\
Llama 4 Maverick & 0.00 & 76.4\% & 0.011 \\
\bottomrule
\end{tabular}
\caption{Agent LLM comparison (skills + source code, best of 2--3 runs; peak accuracy is the test score of the eval-selected configuration). A sharp capability gap separates the models we tested; the weaker two fail to improve upon the baseline.}
\label{tab:model_comparison}
\end{table}

Table~\ref{tab:model_comparison} reveals a sharp division among the models we tested. Sonnet, Opus, and Kimi all produce meaningful optimization trajectories with substantial accuracy gains. On the other side of the gap, Haiku 4.5 barely improves (AUC 0.38) and Llama 4 Maverick fails entirely (AUC 0.00), never recovering from early missteps. The separation is large relative to run-to-run noise: the accuracy difference between the best and worst model is 10.9 pp, roughly twice the 5.4 pp minimum credible improvement we derive below, and the AUC values on either side of the gap differ by an order of magnitude. The two failing models fail in different ways (one cannot execute the optimization protocol at all, the other executes it faithfully but cannot find improving edits), which suggests the binding constraint is model capability rather than any single missing skill or infrastructure issue. The optimization task requires simultaneously diagnosing complex error patterns across multiple fields, generating syntactically valid prompt edits, and reasoning about trade-offs between accuracy and cost. Models that lack the capacity to perform all three of these in concert produce degenerate trajectories. Additionally, Sonnet and Opus benefit from 1-million-token context windows, which reduce compaction frequency and preserve better continuity between iterations. We describe this as a gap among the five models we tested rather than a universal threshold; with only five models, and a frontier that moves quickly, we cannot locate a precise capability boundary or rule out that a smaller model with different training would succeed.

\noindent\fbox{\parbox{0.95\columnwidth}{\textbf{Finding 2.} A sharp capability gap separates the agent LLMs we tested: the weaker two cannot optimize at all (AUC $\leq$ 0.38), while the stronger three produce consistent improvements well outside run-to-run noise. The optimization task requires joint diagnosis, generation, and multi-step reasoning capabilities, and a model lacking them fails outright rather than optimizing more slowly.}}

\subsection{Domain Skills vs. Source Code Access}

\begin{table}[t]
\centering
\small
\begin{tabular}{llrr}
\toprule
\textbf{Agent LLM} & \textbf{Condition} & \textbf{AUC-50} & \textbf{Peak Acc.} \\
\midrule
Sonnet 4.6 & Skills + source & 5.28 & 88.5\% \\
Sonnet 4.6 & Skills only & 5.14 & 87.3\% \\
Sonnet 4.6 & Neither & 4.95 & 87.0\% \\
Sonnet 4.6 & Source only & \textbf{3.05} & 84.7\% \\
\midrule
Opus 4.7 & Skills only & 4.10 & 85.6\% \\
Opus 4.7 & Skills + source & 3.89 & 84.9\% \\
Opus 4.7 & Neither & 3.83 & 84.5\% \\
Opus 4.7 & Source only & 3.16 & 86.2\% \\
\bottomrule
\end{tabular}
\caption{Domain knowledge ablation (best of 2--3 runs; peak accuracy is the test score of the eval-selected configuration). Conditions with skills rank above source-only for both models. Differences between adjacent rows are small and we do not treat them as separable.}
\label{tab:skills_ablation}
\end{table}

Across both models, the conditions with skills occupy the top of Table~\ref{tab:skills_ablation} and source-only sits at the bottom, and the pattern is most pronounced for Sonnet 4.6, where providing source code read access \textit{without} skills lowers AUC from 4.95 (no knowledge at all) to 3.05 (source only). Individually these accuracy differences are small --- the largest gap between any two conditions is 3.8 pp --- so we treat the ordering as a tendency rather than a precise ranking, and we do not claim that any single pair of adjacent rows is separable.

The mechanism behind the tendency is visible in the optimization logs, and is more informative than the accuracy ordering. Source code is voluminous and unstructured; without skills to direct attention, the agent spends iterations reading implementation details and exploring dead-end options instead of making targeted improvements. Agents without skills issue far more exploratory source-code searches per run than agents with skills, and they compact their context correspondingly more often, spending budget on orientation rather than on edits. Skills act as an attention filter, indicating what patterns to look for and what to do about them. For Opus the effect is smaller, likely because stronger models filter irrelevant information more effectively on their own.

\noindent\fbox{\parbox{0.95\columnwidth}{\textbf{Finding 3.} Curated domain skills tend to outperform raw source code read access as a means of supplying domain expertise. We do not observe unstructured source access improving on a no-knowledge baseline, and the optimization logs show it diverting the agent's budget toward orientation rather than targeted edits, suggesting that structured expertise matters more than volume of information.}}

\subsection{Practical Considerations}

\paragraph{Run-to-run variance and a threshold for credible improvement.} With identical starting conditions, peak accuracy varies with a standard deviation of 1.4--3.6 percentage points across runs. Pooling across 44 runs spanning eight ablation conditions gives 2.7 pp, which we use to set a \textbf{minimum credible improvement of roughly 5.4 pp} (two pooled standard deviations) and apply to our own results. The capability gap of Finding 2 clears it comfortably (10.9 pp), as do the individual configuration changes driving most gains; the knowledge conditions of Finding 3 do not, which is why we present that ordering as a tendency. Appendix~\ref{sec:app_variance} gives the per-condition figures.

\paragraph{Context management.} A 50-iteration trajectory generates far more tokens than any model can attend to at once, so a hook compacts the context window whenever token usage exceeds a threshold percentage of it. We use 50\% throughout; Appendix~\ref{sec:app_ctxmgmt} reports the threshold sweep and Appendix~\ref{sec:context_detail} the mechanism.

\subsection{Generalization Across IDP Tasks}
\label{sec:generalization}

\begin{table}[t]
\centering
\small
\begin{tabular}{llrr}
\toprule
\textbf{Task} & \textbf{System} & \textbf{Metric} & \textbf{\$/page} \\
\midrule
\multirow{2}{*}{Extraction} & Human & 81.6\% & 0.102 \\
 & Agent & \textbf{88.0\%} & \textbf{0.022} \\
\midrule
\multirow{2}{*}{Classification} & Human & 100\% & 0.0052 \\
 & Agent & 100\% & \textbf{0.0041} \\
\midrule
\multirow{2}{*}{Packet Split} & Human & 75.9\% & 0.0052 \\
 & Agent & \textbf{84.1\%} & 0.0052 \\
\bottomrule
\end{tabular}
\caption{Agent vs. human expert across three IDP tasks (agent: Sonnet 4.6, skills + source code). Agent matches or beats the expert on all tasks at equal-or-lower cost.}
\label{tab:task_generalization}
\end{table}

Table~\ref{tab:task_generalization} confirms that the approach generalizes beyond the extraction task. On classification (9 classes, 293 documents from the OCR-Benchmark \cite{omniai2025ocrbenchmarkresults}), the agent matches human expert accuracy at 21\% lower cost in 4 iterations, by discovering that OCR is unnecessary for multimodal classification with detailed class descriptions. On packet splitting (15 classes, 500 documents from DocSplit-Poly-Seq \cite{islam2026docsplit}), the agent achieves a 11\% relative improvement over the human expert at identical cost, in 7 iterations.

Before running any of the experiments reported in this paper, in a separate exploratory run, the agent discovered that document class information was encoded in file names and wrote a regex-based strategy that bypassed LLM inference entirely, reaching 100\% accuracy at near-zero cost. We treated this as objective leakage rather than a result: we anonymized every file name to a UUID before conducting the experiments reported here, so no filename signal remains for the agent to exploit. No metrics anywhere in this paper were impacted by this event. We retain the incident as a reward-hacking case study because it illustrates a general property worth designing around: the agent optimizes the stated objective using any available mechanism in the configuration space, and will reach for structural, rule-based, or metadata-driven strategies when they score well.

% Comprehensive Related Work. Merges the former condensed main-paper
% section with the former "Extended Related Work" appendix. Every work
% cited by either version is cited here; the three thematic \paragraph
% groupings and the prose are those of the extended version, with the
% condensed version's distinguishing claims folded into the closing
% sentence of each group.
\section{Related Work}
\label{sec:related}

Our work builds on three lines of research.

\paragraph{Intelligent Document Processing.}
IDP has evolved from template-based systems to LLM-powered pipelines combining OCR, layout analysis, and generative models \cite{shi2024docparsing, chen2025uniparser, kim2022donut}. Recent work addresses specific pipeline challenges: flexible extraction architectures \cite{yoon-etal-2024-lofi}, multi-document packet splitting \cite{islam2026docsplit}, hybrid OCR-LLM strategies for enterprise-scale throughput \cite{wang2025hybridocr}, adaptive agents for regulatory documents \cite{abdellaif2024agenticie}, unsupervised template discovery via multimodal embeddings \cite{sampaio2025clustering}, and evaluation benchmarks for agentic parsing \cite{wu2025parsebench}. While these works advance individual pipeline components (extraction, classification, splitting) or adaptation methods, none address the joint optimization of an entire IDP pipeline configuration (prompts, models, OCR settings, schemas, and parameters), which we perform with our solution.

\paragraph{LLM Agents for System Optimization.}
A growing body of work applies LLM agents to closed-loop optimization. OPRO \cite{yang2024opro} formalizes the pattern as iterative solution generation conditioned on past (solution, score) pairs; TextGrad \cite{yuksekgonul2024textgrad} generalizes it by backpropagating textual feedback through compound AI system graphs. This pattern has been applied to database tuning \cite{zhou2023dbot, lin2025storagextuner}, AutoML pipeline search via MCTS \cite{chi2024sela}, prompt optimization \cite{wang2024promptagent}, agentic workflow generation \cite{zhang2024aflow}, hardware kernel optimization with experience memory \cite{zhang2025accelopt}, and autonomous LLM fine-tuning \cite{li2026ftdojo}. Each of these operates over a single modality (numeric knobs, code, or prompt text) rather than the mixed space of prompts, models, schemas, and structural choices that IDP requires. Our work shares this closed-loop pattern but operates over a configuration space that is broader than any single prior work, requiring simultaneous modification of natural language prompts, categorical model choices, continuous parameters, structural pipeline decisions, and JSON schemas.

\paragraph{Domain Knowledge in LLM Agents.}
Injecting structured expert knowledge into agents improves both accuracy and efficiency. DSPy \cite{khattab2024dspy} encodes task structure as compilable pipeline specifications. SkillForge \cite{chen2026skillforge} iteratively refines domain skills for cloud technical support through failure diagnosis. SkillOS \cite{wang2026skillos} manages skills as structured files in an OS-like repository, achieving performance gains with fewer interaction steps. Our domain skills follow a similar paradigm: self-contained knowledge modules that the agent retrieves on demand for specific IDP failure patterns, but authored by human experts from production engagements rather than automatically generated; we show that human-authored skills from production experience provide more reliable value.

\section{Limitations and Broader Impact}
\label{sec:limitations}

\subsection{Limitations}

Run-to-run variance remains substantial, requiring multiple optimization runs to reliably find strong configurations. Pooled across 44 runs the standard deviation of peak accuracy is 2.7 pp, which means differences below roughly 5.4 pp are not credibly separable from noise; several of our ablation comparisons fall below that bar and we present them as tendencies rather than rankings. This increases the total cost of using the system and means a single run cannot guarantee a near-optimal result.

Domain skills require deliberate authoring effort by experienced practitioners, on the order of an hour per skill for an expert starting cold. Teams applying this approach to a new domain would need to invest in creating analogous knowledge, or accept whatever performance the skills-free condition affords them in that domain.

Our experiments use a fixed 25-document eval set. This size was a deliberate choice rather than a convenience: customers often have limited labeled data available at the start of an engagement, and labeling is expensive enough that requiring hundreds of documents would defeat the purpose of the system. We have not systematically studied the effect of eval set size on optimization quality other than validating that such a relatively small eval set is sufficient for optimization. We therefore cannot say how the method behaves with a substantially larger or smaller eval set, nor where the point of diminishing returns lies.

Our optimization trajectories show no evidence of convergence at 50 iterations, suggesting that longer runs could yield further accuracy gains. We leave extending time horizons to hundreds of iterations to future work.

\subsection{Broader Impact}

\paragraph{Generality of the method:}
Although we instantiate \sysname for document processing, its design is not
specific to documents. The approach applies to any compound system that exposes
a configurable pipeline, a scoring function over a small labeled set, and an
interface for reading errors and applying edits. The heterogeneous
configuration space we optimize, spanning natural-language prompts, categorical
model choices, continuous parameters, structural decisions, and schemas, is
shared by most modern AI systems. Plausible targets include retrieval-augmented
generation, multi-agent workflows, and AutoML pipelines, where prior agentic
optimizers typically operate over a single modality; \sysname instead optimizes
the full mixed space jointly. We have not evaluated the approach in these
settings, and we expect the transfer to depend on how readily failures can be
attributed to specific components: IDP supplies a per-field diagnostic signal
that tells the agent precisely what went wrong and where, whereas in retrieval
or multi-agent settings a poor final answer is harder to trace to the
configuration choice responsible for it.

\paragraph{Transferable findings and assets:}
Our empirical results offer guidance beyond IDP. The capability threshold
indicates that the optimizing model should not be economized, since a weaker
model fails to optimize rather than merely optimizing more slowly. The tendency
of curated skills to outperform unstructured source-code access points to a
lesson in context engineering: for long-horizon agents, what matters appears to
be structured relevance rather than raw volume of available information. Our
context-management and variance-mitigation
procedures generalize to any agent that runs for extended horizons and exceeds
its context window. Several assets are reusable in isolation, including the
evaluation harness that versions configurations to make exploration reversible,
the append-only log that serves as memory and an audit trail, and the
domain-skill format for encoding expert heuristics.

\paragraph{Accessibility:}
By reducing specialist configuration effort from tens of person-hours to hours
of autonomous operation, this class of system lowers the expertise barrier to
deploying compound AI. We view the effect as augmentation rather than
replacement, shifting expert effort from repetitive per-task tuning toward
encoding reusable knowledge and reviewing outcomes.

\paragraph{Risks and responsible use:}
An agent that optimizes any available signal will exploit unintended ones; in
an exploratory run preceding the experiments reported here, ours discovered a
filename-based shortcut that leaked classification labels
(Section~\ref{sec:generalization}). Objective specifications and evaluation
sets must therefore be audited for leakage before optimization begins, and
identifying metadata anonymized, as we did by replacing every file name with a
UUID. Held-out validation remains essential, as optimizing on a small labeled
set can yield configurations that do not generalize. We note that leakage is
defined relative to the deployment condition rather than in the abstract: if a
customer's file names genuinely encode document class at inference time, then a
filename rule is a legitimate and highly efficient solution for them, and the
failure in our case was that our evaluation set carried a signal the production
setting would not. Because such systems process sensitive data, including healthcare
and financial documents, they should operate within the data owner's boundary
and receive only the labeled samples required, and autonomously generated
configurations should be reviewed before deployment.

\section{Conclusion}

We presented \sysname, a deployed agent that autonomously optimizes IDP pipeline configurations in under two hours, exceeding human expert accuracy at lower cost across extraction, classification, and packet splitting tasks. The architecture is domain-agnostic and applies to any compound processing system with a configurable pipeline, a scoring function, and a small labeled set. Our ablations show a sharp capability gap among the agent LLMs we tested, below which optimization fails outright, that curated skills tend to outperform raw source code read access, and that context management is necessary but robust across a range of settings. We expect these findings to transfer beyond IDP to RAG pipelines, multi-agent workflows, and other enterprise AI systems where configuration bottlenecks deployment at scale, though we have not tested this: such settings may lack the clean per-field diagnostic signal that makes IDP failures straightforward to attribute. Run-to-run variance remains a practical challenge, and domain skills demand upfront authoring effort that may not scale without automation.

Future work includes structured search methods (tree search, evolutionary strategies) as alternatives to free-form LLM reasoning, automated skill generation from optimization logs to reduce authoring cost, and extending time horizons to explore whether accuracy continues improving beyond 50 iterations as our trajectories suggest.

% =====================================================================
% References. aaai2027.sty sets the bibliographystyle (aaai2027.bst)
% itself, so no \bibliographystyle command may appear here. AAAI forbids
% \clearpage, so the references flow directly after the text (the style
% file loads placeins and inserts a float barrier for us).
% =====================================================================
\bibliography{references}

% =====================================================================
% Technical appendix.
%
% Placed after the references, mirroring the AMLC version's layout. AAAI
% permits this only as "Supplementary Material / Technical Appendices",
% and that policy varies by conference -- check the AAAI-27 call. If
% AAAI-27 requires the appendix to be part of the main paper instead,
% move this whole block to just before \bibliography above, where it
% becomes a "Content Appendix" and counts toward the page limit.
% =====================================================================
\clearpage
\newpage
\appendix

\section{Experimental Protocol}

\subsection{Human--Agent Comparison Protocol}
\label{sec:app_parity}

The comparison is matched on access: the human experts and the agent operated over the same labeled documents, the same set of available models and OCR backends, and the same configuration space, and both were scored by the same evaluation harness on the same held-out test set. The specialists were optimizing for accuracy, so 81.6\% represents their best effort rather than a configuration abandoned once a business target was met. There is one intentional asymmetry: the agent was subject to a \$0.05/page cost constraint, whereas the human baseline was uncapped and its final configuration costs \$0.102/page. The agent therefore reached higher accuracy while confined to less than half of the human's per-page cost budget. The two sides differ in effort rather than access: weeks of specialist time against a few hours of unattended computation.

\subsection{Test-Set Isolation Protocol}
\label{sec:app_isolation}

The test set is inaccessible to the optimizer by construction, not merely by convention. The test documents are never uploaded to the AWS account in which the agent runs, so no tool available to the agent can read them or score against them. Every configuration the agent selects, and every decision it makes, is based solely on eval-set accuracy. Test-set inference is performed post hoc, after optimization has terminated, in a separate account: we take the configurations selected by eval accuracy and run them against the test set purely to measure generalization and to draw Figure~\ref{fig:convergence}. Consequently the test curves we report are diagnostic rather than selective, and no reported number is obtained by maximizing over test performance.

\subsection{Cost Accounting}
\label{sec:app_cost}

The 4.6$\times$ figure compares \textit{final per-page inference cost}, computed with the same formula for the human and agent configurations (\$0.102/page vs.\ \$0.022/page). It excludes the cost of optimization itself, which we report separately here. The run shown in Figure~\ref{fig:convergence} cost \$139.47 in agent and evaluation inference. This is a one-time cost per document type, incurred once and amortized over the deployment's lifetime, and it should be weighed against the corresponding one-time cost on the human side --- the salaries paid for weeks of specialist and domain-expert time --- which is larger by orders of magnitude.

\subsection{Run-to-Run Variance and the Credible-Improvement Threshold}
\label{sec:app_variance}

With identical starting conditions, peak accuracy varies with a standard deviation of 1.4--3.6 percentage points across runs, because the agent explores different configuration regions depending on early stochastic choices. Skills reduce this variance (0.9 pp with Sonnet + skills-only across 4 runs). Pooling across 44 runs spanning eight ablation conditions gives a standard deviation of 2.7 pp, which we use to set a \textbf{minimum credible improvement of roughly 5.4 pp} (two pooled standard deviations). We apply this threshold to our own results. The capability gap of Finding 2 clears it comfortably (10.9 pp between the best and worst model), as do the individual configuration changes that drive most of the gains, such as adding field descriptions (+29.9 pp) and enabling layout-aware OCR (+12.6 pp). The differences among the knowledge conditions of Finding 3 do not clear it, which is why we present that ordering as a tendency. Running $k$ times and selecting the best by eval accuracy is necessary in practice; this multiplies the one-time optimization cost by $k$.

\section{Implementation Details}

\subsection{Context-Management Threshold}
\label{sec:app_ctxmgmt}

A 50-iteration trajectory generates far more tokens than any model can attend to at once. Before every model call, a hook checks token usage against a configurable threshold percentage of the context window. When exceeded, the context window is compacted (details in Appendix~\ref{sec:context_detail}). At a 50\% threshold, compaction fires every $\sim$18 iterations with minimal impact on accuracy. At 10\%, it fires $\sim$5 times \textit{per} iteration with substantial negative impact on accuracy. We use 50\% throughout.

\subsection{Deployment Infrastructure}
\label{sec:app_infra}

The system runs on Amazon Bedrock AgentCore, deployed via CloudFormation with DynamoDB state persistence, S3 artifact storage, and a web frontend that allows non-technical users to launch and monitor asynchronous multi-hour optimization runs.

\subsection{Domain Skill Examples}
\label{sec:app_skills}

For example, \textit{token-limit-fix} is triggered when documents produce truncated JSON at 0\% accuracy and prescribes increasing the output token limit or decomposing the schema; \textit{visual-spatial-extraction} is triggered when fields depend on spatial position (checkboxes, tables) and prescribes enabling layout-aware OCR features; \textit{multilingual-documents} is triggered for non-Latin scripts and prescribes switching to a backend with broader language coverage.

\section{Context Management Details}
\label{sec:context_detail}

A 50-iteration optimization trajectory accumulates detailed metrics, tool outputs, and source document images, far exceeding what any model can attend to simultaneously. Before every model call, a hook checks current token usage against a configurable threshold expressed as a percentage of the model's context window. When exceeded, the system: (1) summarizes older conversation history via a separate LLM call, producing structured bullet-point summaries of past decisions, tool results, and key findings; (2) preserves the most recent messages intact, maintaining immediate working context; and (3) re-injects the full optimization log, ensuring the agent retains its complete decision history regardless of how many compaction cycles have occurred. The threshold percentage trades information retention against context budget. In our experiments, all runs would have exceeded the 1M-token context window without compaction. At a 50\% threshold, compaction occurs roughly every 18 iterations with minimal accuracy impact. At 10\%, it fires approximately 5 times per iteration, consuming the time budget on summarization; affected runs complete fewer than half their allotted iterations within the 8-hour timeout. Thresholds of 30--60\% are practical; we use 50\% throughout.

\section{Example Domain Skill}
\label{sec:example_skills}

Each domain skill is a structured markdown document retrieved by the agent on demand. Below we present one representative skill (simplified for exposition) that illustrates the format. Skills begin with a YAML header specifying the skill name and a short description used for retrieval, followed by sections defining when the skill applies, how to diagnose the issue, and what configuration changes to make.

\subsection{Skill Header}

% Rewrapped to fit the 3.3in AAAI column. \footnotesize is explicitly
% sanctioned for monospaced material by the AuthorKit template.
\begin{quote}
\footnotesize
\begin{verbatim}
---
name: visual-spatial-extraction-
  challenges
description: Document known challenges
  with fields requiring complex visual
  parsing (checkboxes, damage diagrams,
  spatial alignment) and provide
  mitigation strategies within the
  current architecture.
---
\end{verbatim}
\end{quote}

\noindent The header's \texttt{description} field is used by the agent's skill retrieval mechanism to match observed error patterns to relevant skills.

\subsection{Skill: Visual-Spatial Extraction Challenges}

\paragraph{Trigger:} Specific fields have very low accuracy ($<$60\%) while overall extraction performs well. The failing fields involve visual elements: checkboxes, spatial table alignment, or handwritten annotations.

\paragraph{Diagnosis:}
\begin{enumerate}
    \item Identify low-accuracy fields from per-field evaluation metrics.
    \item Inspect source documents to confirm the failing fields require spatial reasoning (e.g., distinguishing left-column vs.\ right-column check marks in a grid).
    \item Confirm the model is extracting \textit{wrong} values from adjacent visual elements (spatial confusion) rather than extracting nothing (prompt issue).
\end{enumerate}

\paragraph{Resolution:} Apply one or more mitigation strategies in order of increasing cost:
\begin{enumerate}
    \item \textbf{Spatial prompts:} Add highly specific location instructions to the field description (e.g., ``Look for a check mark to the LEFT of the code value. Do NOT confuse with marks on the RIGHT side of the same row.'').
    \item \textbf{Image resolution:} Increase DPI and target image dimensions to give the model better visual input for spatial parsing.
    \item \textbf{Stronger model:} Switch to a more capable multimodal model for the extraction stage.
    \item \textbf{Exclude and flag:} If accuracy remains unacceptable, set the field's evaluation weight to zero and recommend human review for that field.
\end{enumerate}

\subsection{Skill Format Summary}

All 27 skills follow this same structure: a YAML header (name + description for retrieval), followed by trigger conditions, diagnostic steps, and ordered resolution strategies. The key design properties are:
\begin{itemize}
    \item \textbf{Self-contained:} Each skill can be applied independently without reading other skills.
    \item \textbf{Actionable:} Resolution strategies include specific tool calls and configuration edits, not just general advice.
    \item \textbf{Observable triggers:} Trigger conditions are defined in terms of metrics the agent can directly observe (accuracy scores, error patterns, output format).
    \item \textbf{Ordered strategies:} When multiple fixes exist, they are ordered by cost or invasiveness, allowing the agent to try the least disruptive option first.
\end{itemize}

\subsection{Reproducibility}

The benchmark datasets underlying our reported results are publicly available: RealKIE FCC-Verified for extraction, and the classification and splitting benchmarks cited in Section~\ref{sec:generalization}. The evaluation protocol is specified in Section~\ref{sec:setup}, including the eval/test split and the exact-match scoring criteria, so the accuracy numbers we report are reproducible by anyone with a comparable configurable pipeline.

The domain skills themselves are proprietary. They encode diagnostic knowledge from customer engagements and cannot be released in full, which is why we provide one representative skill above in complete form rather than the whole library. The format is fully specified, and the skills are the kind of artifact a practitioner would author from their own engagement experience rather than reuse verbatim from ours; we consider the format and the authoring process, not the specific 27 documents, to be the transferable contribution. The pipeline being optimized is likewise an internal production system, so the specific configuration space in Section~\ref{sec:config_space} is not reproducible verbatim, though the method requires only that some configurable pipeline expose an equivalent surface.

\section{Qualitative Analysis: Single Run Trace}
\label{sec:qualitative}

We present a detailed trace of one optimization run (Sonnet 4.6 for the agent LLM, skills + source code, 50 iterations, 4h 55m) to illustrate the agent's decision-making process. The agent improved accuracy from 28.5\% to 85.2\% (+56.7 pp) while reducing cost per page by 21\%.

\subsection{Starting and Final Configurations}

\paragraph{Starting configuration:} All field descriptions empty, generic extraction prompt, Claude Sonnet 4.6 as extraction model, no OCR, no few-shot examples. Achieves 28.5\% accuracy at \$0.047/page.

\paragraph{Final configuration (v42):} Claude Haiku 4.5 with Textract LAYOUT OCR, domain-specific system prompt, text-only extraction (no document images), 2 text-only few-shot examples, rich field descriptions with column-label guidance. Achieves 85.2\% at \$0.038/page.

Key differences between starting configuration $\rightarrow$ final configuration:
\begin{itemize}
    \item \textbf{Model:} Sonnet 4.6 $\rightarrow$ Haiku 4.5 (cheaper, sufficient with good prompts)
    \item \textbf{OCR:} Disabled $\rightarrow$ Textract LAYOUT (provides structured text)
    \item \textbf{Modality:} Multimodal (text+image) $\rightarrow$ Text-only (images removed)
    \item \textbf{Descriptions:} Empty $\rightarrow$ Rich with column-label guidance
    \item \textbf{Few-shot:} None $\rightarrow$ 2 complementary text-only examples
    \item \textbf{Prompts:} Generic $\rightarrow$ Domain-specific with verbatim extraction rules
\end{itemize}

\subsection{Optimization Trajectory}

Table~\ref{tab:trajectory} shows key milestones. The run naturally divides into phases: bootstrap (model/OCR selection), field-level refinement, few-shot examples, and a late-stage breakthrough when the agent discovers that removing document images improves accuracy.

\begin{table}[t]
\centering
\small
\begin{tabular}{rrlr}
\toprule
\textbf{Iter} & \textbf{Acc.} & \textbf{Key Change} & \textbf{$\Delta$} \\
\midrule
1 & 28.5\% & Baseline (empty config) & --- \\
2 & 58.4\% & Rich field descriptions & +29.9 \\
4 & 71.0\% & Haiku + LAYOUT OCR & +12.6 \\
15 & 73.9\% & Domain system prompt & +2.9 \\
19 & 74.8\% & 2 few-shot examples & +0.9 \\
35 & 77.2\% & Verbatim extraction rules & +2.4 \\
36 & 81.0\% & Remove document images & +3.8 \\
38 & 83.2\% & Column-label guidance & +2.2 \\
42 & 85.2\% & Description column labels & +2.0 \\
\bottomrule
\end{tabular}
\caption{Key milestones from a single 50-iteration optimization run. $\Delta$ shows improvement over the previous milestone.}
\label{tab:trajectory}
\end{table}

\subsection{Agent Trace: Selected Iterations}

We highlight three iterations that illustrate different aspects of the agent's reasoning.

\paragraph{Iteration 2: Field descriptions (+29.9 pp).}

\begin{quote}
\small
\textit{Observation:} LineItemDays at 15\% --- model extracting ``1'' (spots count) instead of day patterns. LineItems truncated at 10 items on long documents.\\
\textit{Diagnosis:} Empty field descriptions provide no guidance; the model guesses field semantics from names alone.\\
\textit{Action:} Populate all 11 field descriptions with rich context. For LineItemDays: ``7-character MTWTFSS pattern where each position is Y/N or the day letter. Example: MTWTF-{}-''.\\
\textit{Result:} +29.9 pp overall. LineItemStartDate +50 pp, LineItemEndDate +58 pp, LineItemRate +50 pp.
\end{quote}

\paragraph{Iteration 36: Removing document images (+3.8 pp).}

\begin{quote}
\small
\textit{Observation:} Text-only Sonnet reference (v12) achieved 76.2\% vs.\ Haiku multimodal at 77.2\%. OCR text appears sufficient for tabular invoices.\\
\textit{Diagnosis:} Document images may add noise for structured tabular extraction --- OCR text provides cleaner column-aligned signal.\\
\textit{Action:} Remove \texttt{\{DOCUMENT\_IMAGE\}} from task prompt. Model receives only Textract OCR output.\\
\textit{Result:} +3.8 pp overall, false negatives halved (88$\rightarrow$45). The largest single improvement in the run. Cost reduced 13\% (no image tokens).
\end{quote}

\paragraph{Iteration 38: Column-label guidance (+2.2 pp with spillover).}

\begin{quote}
\small
\textit{Observation:} LineItemEndDate regressed after removing images --- model confuses date columns without visual layout cues.\\
\textit{Diagnosis:} In text-only mode, the model needs explicit column-name guidance to identify the correct fields.\\
\textit{Action:} Add to LineItemEndDate description: ``Extract from columns labeled End Date, Thru, Through, To, or End.''\\
\textit{Result:} +2.2 pp overall --- but improvement occurred across \textit{all} fields, not just EndDate. Column-label guidance helped the model build a better mental model of the table structure.
\end{quote}

\noindent For example, the final LineItemEndDate field description (starting from an empty placeholder) reads:

\begin{quote}
\small
\textit{``The end date of the advertising spot airing. Extract from columns labeled `End Date', `Thru', `Through', `To', or `End'. Usually shown in MM/DD/YY format. Return null if the invoice has no separate end date column. If shown, extract exactly as printed.''}
\end{quote}

\subsection{Notable Patterns}

Several patterns were observed consistently throughout this run:
\begin{itemize}
    \item \textbf{Null instructions cause cascade regression:} Any instruction telling the model ``return null when X'' caused Haiku to over-generalize, leading to false positive increases across unrelated fields. Confirmed 6 times across iterations.
    \item \textbf{Two few-shot examples is optimal:} Three examples always caused regression regardless of content. Two complementary examples (one with populated fields, one showing null patterns) was the consistent optimum.
    \item \textbf{Text-only examples suffice:} Removing images from few-shot examples matched or improved accuracy (+0.26 pp), suggesting structured text demonstrations are sufficient.
    \item \textbf{Changes interact non-locally:} Adding ``verbatim'' to the system prompt unexpectedly halved LineItemDays false positives --- a field the instruction was not targeting.
\end{itemize}

% Check whether AAAI-27 requires a reproducibility checklist to be
% included in the paper. If so, copy ReproducibilityChecklist.tex from
% the AuthorKit into this directory and uncomment the following line.
% \input{ReproducibilityChecklist.tex}

\end{document}